\documentstyle[12pt,aasms4]{article}
\def\kms{km~s$^{-1}$}
\def\etal{{\it et al.}}

\lefthead{M\'endez \etal}
\righthead{Intergalactic planetary nebulae in Virgo}

\begin{document}

\title{More evidence for an intracluster planetary nebulae 
       population in the Virgo cluster}

\author{R. H. M\'endez}
\affil{Munich University Observatory, Scheinerstr.1, 
                                               Munich 81679, Germany}
\authoremail{mendez@usm.uni-muenchen.de}

\author{M. A. Guerrero}
\affil{Instituto de Astrof\'\i sica de Canarias, E-38200 La Laguna, 
                                                     Tenerife, Spain}
\authoremail{mar@ll.iac.es}

\author{K. C. Freeman}
\affil{Mt. Stromlo and Siding Spring Observatories, 
                              Weston Creek P.O., ACT 2611, Australia}
\authoremail{kcf@mso.anu.edu.au}

\author{M. Arnaboldi}
\affil{Osservatorio Astronomico di Capodimonte, 
                              V. Moiariello 16, Napoli 80131, Italy}
\authoremail{magda@cerere.na.astro.it}

\author{R. P. Kudritzki and U. Hopp}
\affil{Munich University Observatory, Scheinerstr.1, 
                                               Munich 81679, Germany}
\authoremail{kudritzki@usm.uni-muenchen.de, hopp@usm.uni-muenchen.de}

\author{M. Capaccioli}
\affil{Osservatorio Astronomico di Capodimonte, 
                              V. Moiariello 16, Napoli 80131, Italy}
\authoremail{capaccioli@astrna.na.astro.it}

\and

\author{H. Ford}
\affil{Physics and Astronomy Dept., The John Hopkins University, 
                      Homewood Campus, Baltimore, MD 21218, U.S.A. }
\authoremail{ford@jhufos.pha.jhu.edu}

\begin{abstract}

We surveyed a 50 arcmin$^2$ region in the Virgo cluster core to
search for intergalactic planetary nebulae, and found 11 candidates 
in the surveyed area. The measured fluxes of these unresolved sources
are consistent with these objects being planetary nebulae from an 
intergalactic population of stars, although we cannot exclude some 
minor contamination of our sample by redshifted starburst galaxies.  
We compute the cumulative luminosity function of these 11 planetary 
nebula candidates. If we assume that they belong to the Virgo cluster, 
their cumulative luminosity function is in good agreement with 
planetary nebula luminosity function simulations for a typical 
stellar population of ellipticals or spiral bulges.  This comparison 
allows us to estimate the surface mass density of the intergalactic 
stellar population at the surveyed field in the cluster core.

\end{abstract}


\section{Introduction}

A recent study of the dynamics of the outer regions of NGC~4406 
(M~86) using radial velocities of planetary nebulae (PNs) 
by Arnaboldi \etal \ (1996) produced the serendipitous 
discovery of 3 PNs whose velocities were near the mean redshift 
of the Virgo cluster (around $+1400$ \kms). Since NGC~4406 has a 
peculiar redshift of about $-230$ \kms, this led to the conclusion 
that the 3 PNs do not belong to NGC 4406, but are members of
an intergalactic or intracluster diffuse stellar population in the
Virgo cluster. More recently, other independent observations
gave further evidence for intracluster stars: a deep HST image 
of a blank field near M 87 revealed several hundreds of red
giants (Ferguson \etal \ 1996, 1997), and Theuns and Warren (1997) 
reported the detection of several intracluster PN candidates
in the Fornax cluster.

The existence of a diffuse intracluster stellar population has been 
expected, as a consequence of the stripping of stars from galaxies 
in clusters through the effect of fast encounters with other galaxies 
and the interaction of the galaxies with the tidal field of the
cluster itself. This process (harassment) was simulated 
recently by Moore \etal \ (1996).  Planetary nebulae are 
ideal tracers of such an intracluster stellar population.  
They are relatively easy to find and, from surveys of several 
fields in the Virgo Cluster, it is possible to map the surface 
density distribution of the intracluster population,
compare it with the distribution of the galaxies in the cluster 
(see Binggeli \etal\ 1987), and estimate its total mass. In addition, 
the fact that the detected flux is concentrated in one spectral line
([O III] $\lambda5007$) makes it relatively easy to measure the 
radial velocities of these intracluster PNs with adequate accuracy. 
Using the PN radial velocities, we can compare the kinematics of 
this intracluster population with the kinematics of the
diffuse population produced in simulations of galaxy harassment. 

As a by-product of the PN detections, the luminosity function of 
the intracluster PNs can be used to obtain a new estimate of the
distance to the Virgo cluster, less affected by problems of
incompleteness of galactic PN samples and the back-to-front
uncertainties related to the galaxy morphological types and 
their spatial segregation in the cluster (e.g. Jacoby \etal \ 
1990). Given a large enough sample of intracluster PNs it will
be interesting to compare them with the Jacoby \etal \ (1990) PNs,
because the intracluster sample may have slightly brighter PNs;
across the cluster depth we expect to detect preferentially those
PNs which are closer to us (Jacoby 1997).

To enlarge the statistics on the intracluster PN population in 
the Virgo cluster core, we decided to survey an additional field 
in Virgo, using the well-established [O III] on-band/off-band 
technique of Jacoby \etal\ (1990). We selected a field near the 
core of the Virgo cluster but well away from bright galaxies and 
bright stars, and made a deep search for intracluster PNs.  In 
this letter we present the discovery of 11 PN candidates 
(Sections~2 and 3), and discuss of possible contaminations from 
starburst galaxies at high redshifts in Section~4. We discuss the 
cumulative PN luminosity function (PNLF) and the surface mass 
density of the diffuse population in Sections~5 and 6.
Conclusions are given in Section~7.

\section{Observations and reductions}        % This is Section 2.

The highest priority field for our survey was centered at 
RA(2000)$ = 12^h 26^m 32\fs1$, 
$\delta(2000) = +12\arcdeg 14\arcmin 39\arcsec$, in the core
region, near the center of group A in the Virgo cluster (see 
the isopleths in Fig. 4 of Binggeli \etal \ 1987), where there
are no bright stars and no giant galaxies, in order to survey
only the diffuse intracluster population. Using the NED 
database\footnote{The NASA/IPAC Extragalactic Database (NED) is
operated by the Jet Propulsion Laboratory, California Institute of
Technology, under contract with the NASA.} we verified that there
is no known dwarf or low surface brightness galaxy belonging to the
Virgo cluster within the selected field. The only previously known 
galaxy in the field is number 166 in Drinkwater \etal \ (1996), a 
background object at a redshift $z = 0.0845$.
We also verified that an extrapolation of the halos of nearby giant 
galaxies (M 84, M 86, M 87) based on the r$^{1/4}$ law, using data
from Peletier \etal \ (1990) and Caon \etal \ (1990), gives such 
a low blue surface brightness in our selected field (fainter than 
31 mag arcsec$^{-2}$) that no detection of PNs belonging to those
galaxies can be expected. For comparison, in Section 6 we derive 
a blue surface brightness of 28.3 mag arcsec$^{-2}$ for the 
detected stellar intracluster population.

The observations were made with the Prime Focus CCD camera
of the 4.2m William Herschel telescope (Observatorio del Roque
de los Muchachos, La Palma, Spain) on the 5-8 of May, 1997.
We used a 2k $\times$ 2k Loral CCD with 15 $\mu$m pixels 
(0.26 arcsec pixel$^{-1}$) and $95\%$ quantum efficiency at 
5000 \AA. The on-band and off-band interference filters were  
provided by the Instituto de Astrof\'\i sica de Canarias (IAC 
filters numbers 7 and 18): their central wavelengths were 5028 
\AA\ and 5457 \AA, maximum transmissions 87\% and 77\%, 
equivalent widths 45.2 and 147.7 \AA, and FWHMs 50 and 200 \AA\ 
respectively.  Since the filters were located in the f/2.8 
converging beam, and the operating temperature was about
12\arcdeg C, the central wavelengths were blueshifted to 5021 
\AA\ and 5450 \AA, for the on- and off-band filter. For the
mean redshift of the Virgo cluster, and the velocity dispersion
$\sigma_{\rm cluster} = 800$ \kms\, the width of the on-band filter
covered all the velocities in the $-2.6 \sigma_{\rm cluster}$ to 
$+1.2 \sigma _{\rm cluster}$ range, centered at the mean redshift of 
the Virgo cluster.  The interference filters were smaller than the 
normal filters used at the WHT Prime Focus camera, and the effective 
surveyed field was circular with a radius of 4 arcmin. We had 
sporadic cirrus clouds during the first two nights; the third night 
was photometric. The typical seeing was 0.9 arcsec, and 1.1 arc sec 
at large airmasses.

We were able to make $10 \times 3600$ s exposures through 
the on-band filter, with air masses less than 1.4, and 
$8 \times 1500$ s exposures through the off-band filter 
(frequently at higher air masses).  We reached a fainter
limiting magnitude in the off-band images.  We also obtained 
on-band and off-band exposures of the standard star G138-31 
(Oke 1990) for the flux calibrations, and several twilight 
flat fields with high number counts.

The CCD reductions were made using standard tasks
provided by IRAF\footnote{IRAF is distributed by the National 
Optical Astronomical Observatories, operated by the Association 
of Universities for Research in Astronomy, Inc., under contract 
to the National Science Foundation of the U.S.A.}. After
bias-level subtraction and flatfielding, the on-band images 
were registered and combined, to minimize the effect of bad 
columns (the telescope was dithered between exposures) and 
eliminate cosmic ray events. The image combination was made 
with the IRAF task COMBINE, using the AVERAGE option and CCDCLIP
rejection (pixel rejection based on CCD noise parameters).

\section{Detections and photometry}    % This is Section 3.

The resulting combined on- and off-band images were split into 
subimages, and each of the 16 on-band subimages was carefully 
inspected for PN [O III] emission by blinking it with the 
corresponding off-band subimage. A test image obtained by 
dividing the on-band image by the off-band image was 
also blinked with the off-band image.  With these two blinking 
procedures, several candidates were selected. For each candidate 
we inspected the individual images, to check whether the apparent 
emission was an artifact resulting from the bad CCD columns or 
from an uneliminated partial superposition of cosmic ray events 
(a few such cases were found). After careful examination, we 
identified 11 PN candidates which satisfied the following 
criteria:

\noindent (1) they were clearly visible (detection above 2.5 sigma)
              in the combined on-band image, but not visible (below
              1 sigma) in the combined off-band image,

\noindent (2) of similar brightness 
              in all of the individual on-band images,

\noindent (3) unresolved point-like sources.

Two additional unresolved sources were found: although they appear 
much brighter in the on-band image, they are also visible in the 
off-band image. We will discuss them below in more detail.

Arnaboldi \etal\ (1996) found 3 intergalactic PNs in an area 
of 16 arcmin$^2$, which would imply the presence of 9 PNs in
our 50 arcmin$^2$ field, if we assume a similar PN surface 
density and the same limiting magnitude. Therefore we find good 
agreement between the detected candidates and the expected number 
of intracluster PNs from the previous work. This agreement is 
relevant in particular because the 3 intergalactic PNs of 
Arnaboldi \etal \ are spectroscopically confirmed.

The photometry was made with the PHOT and DAOPHOT routines
in IRAF, following standard procedures: aperture photometry for 
the standard star (an aperture diameter of 14 pixels was adopted)
and PSF-fitting for all point sources in the on-band and off-band 
combined images. The FWHM of the PSF was 4 and 5 pixels for the 
combined on-band and off-band images, respectively; the larger 
FWHM for the off-band image was expected because of the larger 
air masses. A set of 20 stars was used 
(1) to calculate an aperture correction of $-0.23$ mag to the 
instrumental magnitudes, and 
(2) to correct the instrumental magnitudes from the first
two (non-photometric) nights to the system of the images from the
last (photometric) night. This correction was $-0.14$ mag. 

We obtained the fluxes $I_{5007}$ in erg cm$^{-2}$ s$^{-1}$,
using the measurements of the standard star G138-31 and the 
properties of the on-band filter. The $I_{5007}$ flux can be 
expressed in magnitudes $m_{5007}$, 

$$   m_{5007} = -2.5 \ {\rm log}(I_{5007}) - 13.74    \eqno(1)   $$

\noindent (Jacoby 1989). The $m_{5007}$ of the 11 PN
candidates are between 26.8 and 28.6. We estimate errors
of 0.05 mag for the brighter objects and 0.2 mag for the 
fainter ones.

\section{Emission-line galaxies?}  % This is Section 4.

Could our objects be unresolved emission-line galaxies whose 
emission lines have been redshifted into the on-band filter
centered at 5021 \AA\ (e.g. [O II] 3727 at $z=0.35$,
or Lyman~$\alpha$ at $z=3.1$) ? We can estimate how many
high-redshift emission-line galaxies would lie within our field of 
50 arcmin$^{2}$. From our photometry, an object must be fainter 
than 26th (visual) magnitude to be undetected in our off-band 
combined image. We can use the analysis of Ellis \etal\ (1996) 
and Heyl \etal \ (1997) on high redshift starburst galaxies
as a starting point for our discussion. We assume that 
H$_0$ = 70 km s$^{-1}$ Mpc$^{-1}$ and q$_0$ = 0.5. Then, for 
redshifts between 0.340 and 0.354, from eqs. (6), (7), (8) of 
Ellis \etal, we find that the proposed starburst galaxies must 
be fainter than $M_{b_j} = -16$, and that the effective surveyed 
volume is 228 Mpc$^3$. From Figure 21 of Heyl \etal, we read a 
value of 0.022 galaxies Mpc$^{-3}$ mag$^{-1}$ for the luminosity 
function of starburst galaxies. For our surveyed volume of 228 
Mpc$^3$ we derive a total number of 5 starburst galaxies per mag.
Even if these galaxies were to appear unresolved in the on-band 
image, their [O II] 3727 emission would need to be very strong to 
satisfy our observational constraints.  From our filter 
transmission curves, if these high redshift galaxies have a 
$m_{5007}\sim 28$, and remain undetected in the off-band image, 
then their [O II] 3727 equivalent width has to be 100 \AA. 
Figure 19 of Heyl \etal\ indicates that the [O II] equivalent 
widths of these high redshift galaxies do not usually reach this 
level. Additional evidence comes from studies of emission-line 
galaxies detected by Popescu \etal \ (1996). From unpublished 
spectra it is possible to estimate that only about 5\% of those 
emission-line galaxies can have an [O II] equivalent width of
100 \AA. 

We conclude that none of our PN candidates brighter than 
$m_{5007}\sim 28$ can be explained as starburst galaxies at 
$z=0.35$, perhaps only a few of the fainter ones. From similar 
arguments (see also Theuns and Warren 1997) we can also exclude 
galaxies or quasars at higher redshifts as an explanation for the 
majority of our PN candidates. 

Some level of contamination remains possible in our sample, 
and we plan to acquire spectra of our PN candidates in the 
near future. It seems likely that the two point-like sources
reported above, which were visible in the off-band image and 
brighter in the on-band, are starburst galaxies or quasars; 
this also requires spectroscopic confirmation.

\section{The cumulative PN luminosity function} % This is Section 5.

In what follows we will assume that the 11 candidates are 
intergalactic PNs in the Virgo cluster. The sample is limited,
so we cannot build the PN luminosity function as in Jacoby 
\etal \ (1990). We prefer to use the cumulative PN luminosity 
function, which gives for each value of $m_{5007}$ the total 
number of PNs brighter than $m_{5007}$. 

The result is given in Figure 1, where the cumulative PNLF 
is compared with simulated PNLFs. The simulated PNLFs are
calculated for three different sample sizes (by sample size we 
mean the total number of PNs in the surveyed area, of which 
only the brightest are detected). We use numerical PNLF
simulations like those discussed by M\'endez and Soffner 
(1997). These PNLF simulations reproduce the observed PNLF
of M 31's bulge, a population without recent star formation, 
assumed to be sufficiently similar to the one in the outskirts 
of elliptical galaxies.

To transform the measured $m_{5007}$ into absolute 
magnitudes, we adopt a distance modulus of 30.9 mag and an 
interstellar extinction correction of 0.06 mag (Jacoby \etal \ 
1990, M\'endez \etal \ 1993).
The simulated and observed cumulative PNLFs agree well, which 
supports the identification of our candidates as intracluster 
PNs in the Virgo cluster. We do not have enough objects 
for a reliable distance determination, but the 
good agreement between the observed and simulated  PNLFs is a 
check of consistency with the adopted distance modulus. 

\section{The surface mass density of the diffuse population} 
%This is Section 6.

From our rough evaluation of the PNLF sample size in Figure~1, 
we estimate the surface mass density of the diffuse 
intergalactic population.

If $\dot \xi$ is the specific PN formation rate, in 
PN yr$^{-1}$ L$^{-1}_\odot$ ; $L_{\rm T}$ is the total 
bolometric luminosity 
of the sampled population; and $t_{\rm PN}$ is the lifetime of 
a PN (which we take as 30,000 years), then the PNLF sample size 
$n_{\rm PN}$ is given by the following relation:

$$ n_{\rm PN} = \dot \xi \ L_{\rm T} \ t_{\rm PN}    \eqno(2)   $$

\noindent For a more detailed discussion on the practical use 
of this relation, see M\'endez \etal \ (1993) and Soffner \etal \ 
(1996). 

Here we adopt $n_{\rm PN} = 100$, from Figure~1, and 
$\dot \xi = 4\times10^{-12}$ (a typical value for Virgo 
ellipticals, with an uncertainty of about 50\%; see Jacoby \etal \
1990 and M\'endez \etal \ 1993). We derive a surveyed luminosity 
of $8\times10^8 {\rm L}_{\odot}$. Given the uncertainties in 
$n_{\rm PN}$ and $\dot \xi$, we estimate that our result for 
the surveyed luminosity is uncertain by a factor 2 or 3.

Would this intracluster population be detectable through its 
emitted continuum light? The luminosity derived above 
($8\times10^8 {\rm L}_{\odot}$) corresponds to a blue 
surface brightness 
of 28.3 mag arsec$^{-2}$, which would be difficult to detect 
on the angular scale of the Virgo cluster. We conclude that 
there is no inconsistency at this time between the diffuse 
surface brightness of the Virgo cluster and the presence of 
an intracluster stellar population as seen through the PNs.

Binggeli \etal \ (1987) estimate that the central surface 
luminosity density in the Virgo cluster is 
10$^{11} {\rm L}_{\odot} {\rm deg}^{-2}$. This is equivalent 
to $1.4\times10^9 {\rm L}_{\odot}$ in our surveyed area of 50 
arcmin$^2$. Therefore, assuming similar distributions, the 
luminosity contributed by the 
intracluster population could be of the order of 50\% of 
the luminosity from all galaxies in the Virgo cluster. 
This number is quite similar to that found by Theuns and 
Warren (1997) in the Fornax cluster and by Bernstein \etal \
(1995, from a study of the diffuse light) in the Coma cluster.

Assume now for the Virgo intracluster stars 
a mass-to-luminosity ratio of 5, typical for the stellar
content of an old population (we do not include here any 
dark matter possibly associated with such a population). Then 
the total surveyed stellar mass is $4\times10^9$ M$_{\odot}$. 
Again, this could amount to about 50\% of the stellar 
mass from all galaxies in the cluster. At a distance of 15 Mpc 
our surveyed area corresponds to $10^9$ pc$^{2}$. Therefore the 
surface mass density of stars is 4 M$_{\odot} {\rm pc}^{-2}$ for 
the intracluster population at the surveyed position. 

Let us compare the intracluster stellar surface mass density 
of 4 ${\rm M}_{\odot} {\rm pc}^{-2}$ with the surface mass density 
of the total gravitational mass in the Virgo cluster at 
the same position in the sky. A rough estimate of the
latter can be obtained as follows. We take recent studies 
of the Virgo cluster structure as derived from X-ray images 
(B\"ohringer \etal \ 1994) and X-ray spectra (Nulsen and 
B\"ohringer 1995). We adopt the distribution of gravitating 
mass given by Nulsen and B\"ohringer (see their Eq. 13 and 
their Fig. 5), extrapolate it out to a distance of 1 Mpc from 
the center of M 87 (which gives a total gravitating mass of 
1.5$\times 10^{14}$ solar masses), recover the mass density (mass 
per unit volume) at each point and integrate along the line of 
sight through the cluster at the position of our field. In this 
way we obtain a total gravitational surface mass density of 
80 M$_{\odot} {\rm pc}^{-2}$. Since the surface L density of
the intracluster population is 
$8\times10^8\times10^{-9} {\rm L}_{\odot} {\rm pc}^{-2}$,
we conclude that the M/L ratio (where M is the total gravitational 
mass) at the position of our search is approximately 100. 

Nulsen and B\"ohringer estimate that the total gravitational mass 
near the center of the cluster is roughly 15 times the mass of the 
hot gas. We obtain a similar factor (20) for the ratio between 
the total gravitational mass and the mass of the intracluster
population. Then, if we assume spherical symmetry and similar 
distributions as a function of distance to the center, the mass 
of the diffuse intracluster stellar population may be of the same 
order of magnitude as the mass of hot gas in the intracluster medium. 

All these estimates are of course extremely uncertain, and 
our main purpose in making them is to show that it would be 
interesting to repeat the search for PNs at several different 
positions across the Virgo cluster in order to map the 
surface mass density of the diffuse intracluster population.
The contribution from this new population cannot account for 
all the dark matter in the Virgo cluster, but it certainly 
increases the ratio of visible to total matter, 
making it harder to reconcile the results of standard 
big bang nucleosynthesis with an $\Omega_0=1$ universe, as 
discussed for example in White \etal \ (1993).

\section{Conclusions}  % This is Section 7.

Our search confirms our earlier discovery of an intracluster 
stellar population in the Virgo Cluster, and gives a further 
hint that this population may represent a significant 
contribution to the total mass of the cluster. The next steps 
in this program are: (1) survey further fields at different 
distances from the center of the cluster, to map the structure 
of this population; (2) acquire spectra of the detected objects, 
to check their identification as PNs and to start the kinematical 
study of this population; (3) study numerical simulations of star 
stripping, scaled to the Virgo Cluster, to investigate the expected 
kinematics of the stripped stellar population and to evaluate the 
likely mass fraction of stripped stars in the cluster.

\acknowledgements

The William Herschel Telescope is operated on the island of La 
Palma by the Isaac Newton Group in the Spanish Observatorio del 
Roque de Los Muchachos of the Instituto de Astrof\'\i sica 
de Canarias. The work reported here has been supported by the
Deutsche Forschungsgemeinschaft through Grant SFB 
(Sonderforschungsbereich) 375. RHM would like to acknowledge
useful discussions with Ralf Bender and Roberto Saglia.

{}

\clearpage
\begin{figure}
\epsscale{0.9}
%\plotone{virf1.ps}
\figcaption{The cumulative PNLF (full line) derived from our sample of
11 PNs, if we assume them to be at the distance of the Virgo cluster, 
and if we adopt a distance modulus of 30.9. For comparison, we show
three simulated PNLFs, which represent typical PN populations like 
those found in early-type Virgo cluster galaxies or in the bulge of 
M 31. These three simulated PNLFs (dashed lines) were calculated for 
a maximum final mass of 0.63 M$_{\odot}$ and sample sizes of 50, 100 
and 150 PNs. For details, see M\'endez and Soffner (1997). From the 
comparison of the observed cumulative PNLF with the simulated ones we 
estimate a sample size of about 100 PNs for our Virgo intergalactic PNs.}
\end{figure}

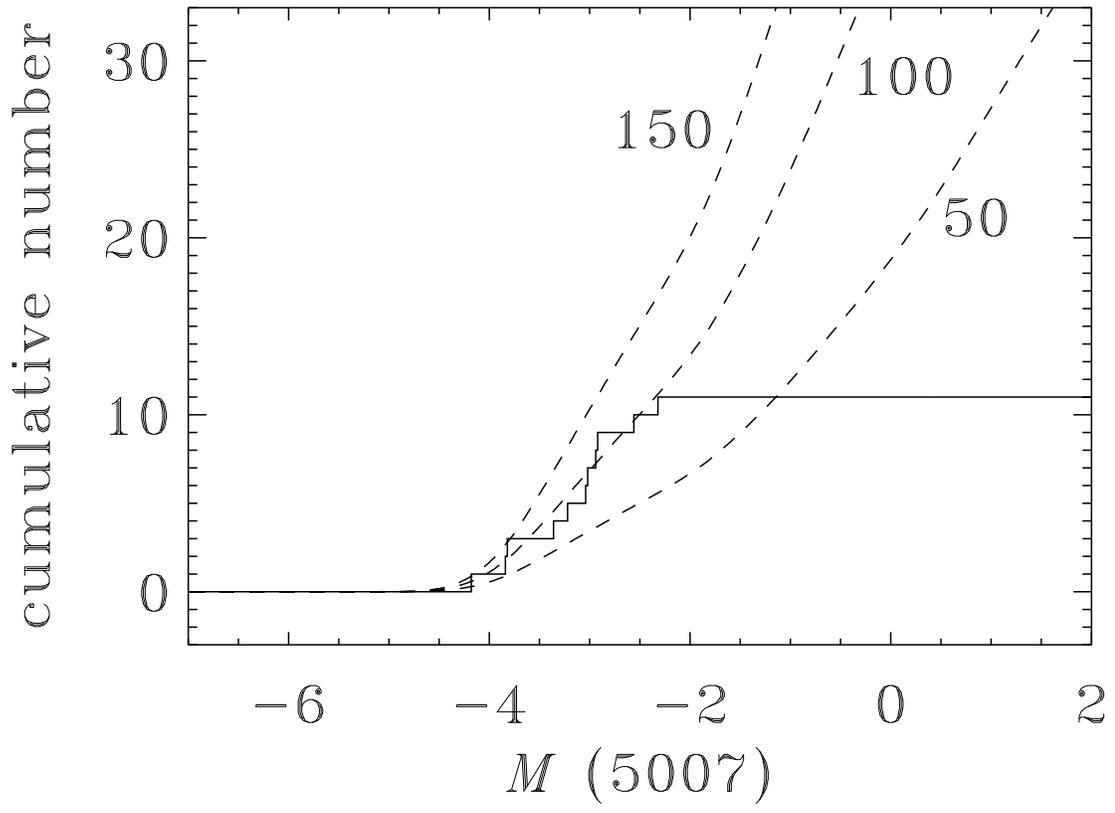
\end{document}